\documentclass[reprint,prb,superscriptaddress,amsfonts,amssymb,amsmath,showpacs,titlepage,longbibliography]{revtex4-1}

\usepackage{hyperref}
\usepackage{latexsym}
\usepackage{verbatim} 
\usepackage{natbib}
\usepackage{graphicx}
\usepackage{textcomp}
\usepackage{xcolor}
\usepackage{amssymb}
\usepackage{amsmath}
\usepackage[english]{babel}
\usepackage{epstopdf}
\usepackage{float}

\begin {document}

\title {Adsorption-induced symmetry reduction of metal-phthalocyanines studied by vibrational spectroscopy}

\author {J. Sforzini}
\affiliation {Peter Gr\"{u}nberg Institut (PGI-3), Forschungszentrum J\"{u}lich, 52425 J\"{u}lich, Germany}
\affiliation{J\"{u}lich Aachen Research Alliance (JARA), Fundamentals of Future Information Technology, 52425 J\"{u}lich, Germany}

\author {F. C. Bocquet}
\email{f.bocquet@fz-juelich.de}
\affiliation {Peter Gr\"{u}nberg Institut (PGI-3), Forschungszentrum J\"{u}lich, 52425 J\"{u}lich, Germany}
\affiliation{J\"{u}lich Aachen Research Alliance (JARA), Fundamentals of Future Information Technology, 52425 J\"{u}lich, Germany}

\author{F. S. Tautz}
\affiliation {Peter Gr\"{u}nberg Institut (PGI-3), Forschungszentrum J\"{u}lich, 52425 J\"{u}lich, Germany}
\affiliation{J\"{u}lich Aachen Research Alliance (JARA), Fundamentals of Future Information Technology, 52425 J\"{u}lich, Germany}

\begin{abstract}
We investigate the vibrational properties of Pt- and Pd-phthalocyanine (PtPc and PdPc) molecules on Ag(111) with high resolution electron energy loss spectroscopy (HREELS). In the monolayer regime, both molecules exhibit long range order. The vibrational spectra prove a flat adsorption geometry. The red shift of vibrational modes and the presence of asymmetric vibrational peaks suggest a moderate interaction of the molecules with the substrate, accompanied by a static charge transfer from the metal to the molecules. The appearance of a particular vibrational mode, which (i) belongs to the $\mathrm{B_{1g}}$ representation of the original fourfold $\mathrm{D_{4h}}$ molecular symmetry group and which (ii) exhibits interfacial dynamical charge transfer (IDCT), proves that a preferential charge transfer from the Ag surface into one of the originally doubly degenerate lowest unoccupied molecular orbitals (LUMOs) of $\mathrm{E_g}$ symmetry takes place, i.e. the electronic degeneracy is lifted and the molecule-surface complex acquires the twofold symmetry group $\mathrm{C_{2v}}$. The vibration-based analysis of orbital degeneracies, as carried out here for PtPc/Ag(111) and PdPc/Ag(111), is not restricted to these cases. It is particularly useful whenever the presence of multiple molecular in-plane orientations at the interface makes the analysis of orbital degeneracies with angle-resolved photoemission spectroscopy difficult.    

\end{abstract}

\maketitle

\section{Introduction}

Organic molecules with $\pi$-conjugated electron systems have been intensively studied in recent years. Apart from a fundamental interest in their electronic properties, this activity is motivated by the wide range of possible applications in the fields of optoelectronics\cite{Friend1999, Blom2007} and spintronics \cite{Sanvito2011}. Among these molecules, metal-phthalocyanines (MPc), i.e. tetrabenzoporphyrazine macrocycles with a metal atom in their center, play an important role, because of their planar geometry, their thermal stability, their suitability for organic molecular beam epitaxy, and their chemical versatility that its brought about by very diverse central metal atoms \cite{Gottfried2015}. In fact, MPc molecules have been employed in organic light emitting diodes\cite{ Blochwitz1998}, field effect transistors\cite{Crone2000,Horowitz1998} and solar cells\cite{Conboy1998}.

The interaction of MPc with metal surfaces is interesting both from a fundamental point of view, in particular regarding the balance between the contributions of the central metal atom and the $\pi$-electron system, and for applications, in which such interfaces are the primary functional elements. For MPc with almost filled $d$-shells, the molecule-substrate interaction is dominated by the chemisorptive or physisorptive interaction of the macrocycle with the metal surface \cite{Gottfried2015}. If there is a chemisorptive contribution to this bond, this often leads to charge transfer from the metal into the lowest unoccupied molecular orbital (LUMO) of the molecule \cite{Gottfried2015}. This is also true in the case of PtPc and PdPc on Ag(111), as we show in this paper.  

An interesting issue in relation to the molecule-substrate interaction in general and the charge transfer in particular is the symmetry of adsorbed MPc molecules. Except for a few cases, e.g., SnPc \cite{Wang2009a}, the central metal fits into the inner cavity and the corresponding isolated MPc belongs to the $\mathrm{D_{4h}}$ symmetry group \cite{Gottfried2015}. Because of the presence of the surface, the symmetry of the molecule is lowered to $\mathrm{C_{4v}}$. However, a further molecular symmetry reduction to a twofold symmetry is often observed in scanning tunneling microscopy (STM) \cite{Chang2008,Karacuban2009, Wang2009a,Wang2009,Cuadrado2010,Snezhkova2015}. 

The symmetry reduction of MPc molecules from fourfold to twofold symmetry raises an intriguing question regarding the charge transfer into the molecule. The LUMO of isolated MPc is twofold degenerate and belongs to the irreducible representation $\mathrm{E_g}$ of the $\mathrm{D_{4h}}$ symmetry group. This degeneracy is closely related to cross-conjugation in the central porphyrazine macrocycle of phthalocyanine molecules \cite{Phelan1968,Gottfried2015}. When charge is transferred into the LUMO, two situations are conceivable: either both orbitals remain degenerate and receive the same amount of charge, or  the degeneracy is lifted and charge is transferred preferentially into one of the two. Degeneracy lifting has been invoked in certain cases to explain the symmetry reduction of MPc molecules \cite{Karacuban2009,Uhlmann2013,Snezhkova2015}, while in other cases equal filling of both  $\mathrm{E_g}$ orbitals was conjectured \cite{Feyer2014}, or the symmetry reduction was assigned to structural effects \cite{Chang2008,Cuadrado2010,Niu2013}.

At first glance, STM seems to be the method of choice for symmetry analysis of molecular adsorbates, because any symmetry breaking should become immediately obvious in the image. However, in STM purely electronic as well as purely geometric effects can both lead to a symmetry reduction in the image, and there are only a few cases in which the origin of the broken symmetry can be determined on basis of STM alone (e.g. Ref.~\onlinecite{Uhlmann2013}). Of course, in most cases geometric and electronic effects will be coupled to each other, because a geometric distortion of the molecule by the influence of the external environment will also break the electronic symmetry within the molecule, and vice versa an electronic symmetry reduction will generally lead to a structural distortion (Jahn-Teller effect \cite{Jahn1937,Kopper2009,Uhlmann2013}). Yet, it is still an important question to ask which is the dominant of the two effects. For example, a strong geometric distortion may lead to a negligible symmetry break of electronic states. 

For this reason, additional experimental methods have been employed to settle the issue of a possible lifting of the degeneracy. These include core level spectroscopy \cite{Snezhkova2015}, near-edge x-ray absorption spectroscopy (NEXAFS) \cite{Borghetti2014}, or density functional theory (DFT) \cite{Snezhkova2015}. Recently, also photoemission tomography \cite{Puschnig2009, Puschnig2011} has been applied to this problem: In the case of FePc/Ag(111) no splitting of the LUMO was found \cite{Feyer2014}, while for CuPc/Ag(110) a clear splitting was observed \cite{Schonauer2016}. Photoemission tomography has the unique advantage that if a degeneracy lifting takes place, the actual orientation of the filled orbital can be determined \cite{Schonauer2016}. However, this orientational sensitivity can also be a drawback, if many molecules with many different orientations are present at the surface. In this case, photoemission tomography results become difficult to analyze \cite{Feyer2014}. 

Here we present an approach that does not suffer from the presence of multiple orientations, because only the internal symmetry of the molecule-surface complex matters. Yet at the same time, if applicable, it leads to unambiguous results regarding electronic symmetry breaking, i.e. degeneracy lifting. This approach is based on vibrational spectroscopy. Specifically, we employ high resolution electron energy loss spectroscopy (HREELS) which in its dipole scattering mechanism is sensitive to infrared (IR) active vibrational modes \cite{Ibach1982}. Due to the surface selection rule, only totally symmetric modes of the molecule-surface complex are IR active \cite{Ibach1982}. Any reduction of the molecular symmetry group upon adsorption may imply that formerly inactive modes can become IR active. Because, as mentioned above, the symmetry reduction can be effected both by geometric and electronic effects, also the corresponding IR activation can have these two distinct origins. However, if additionally the line shape of an activated mode is taken into account, it is (under favorable conditions) possible to unambiguously link the activation of certain vibrational modes to an electronic symmetry breaking. More specifically, we argue here that if a mode that indicates a particular symmetry reduction has a Fano line shape, this proves an electronic contribution (i.e. degeneracy lifting) to this symmetry reduction, because the Fano line shape indicates an interfacial dynamical charge transfer (IDCT), and for an IDCT to be observable in a MPc molecule, there must exist an imbalance in the occupation of the two $\mathrm{E_g}$ LUMOs. Hence, their degeneracy must be broken.

\begin{figure*}
	\centering
		\includegraphics [scale=0.7]{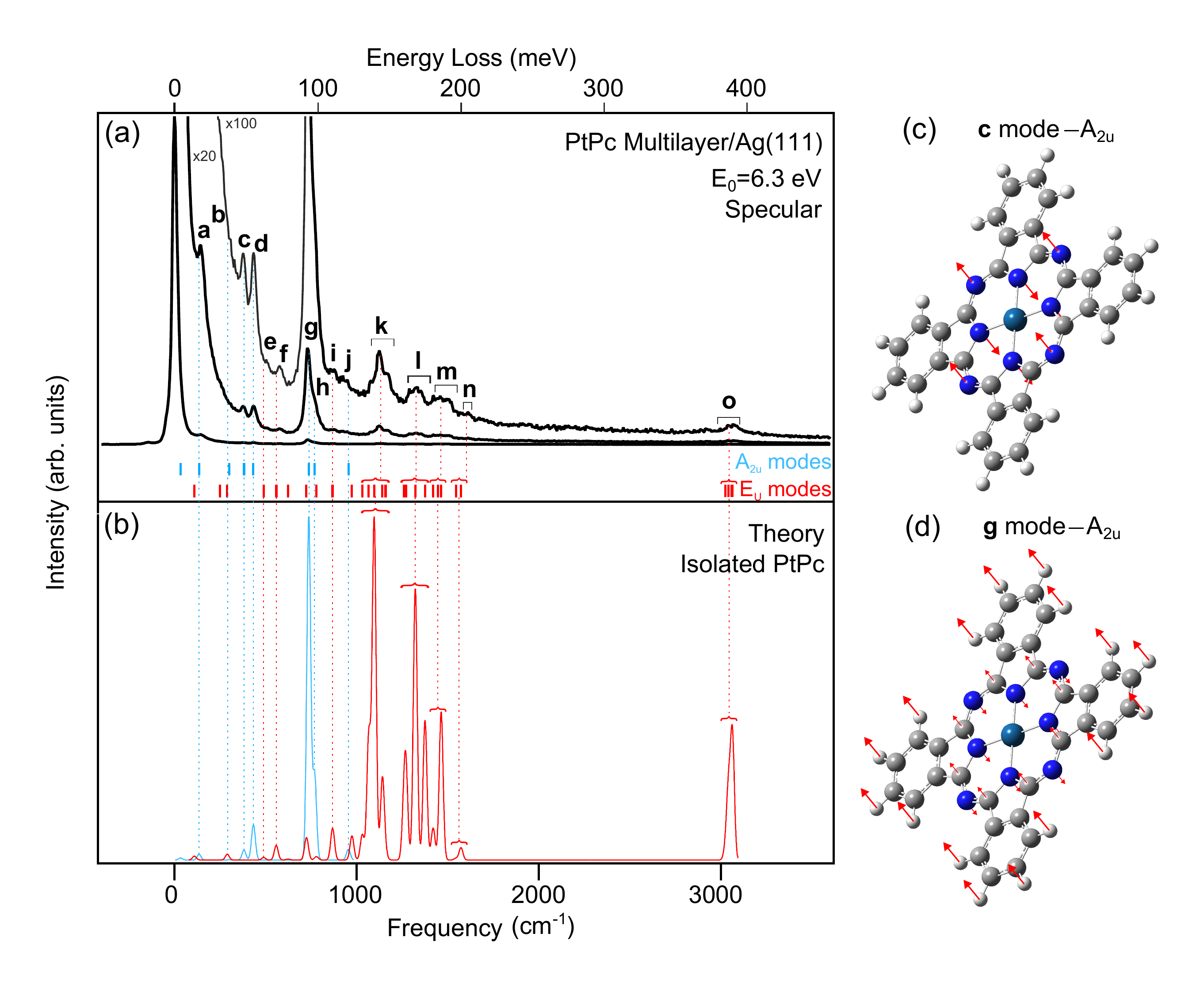}
	\caption{(a) The experimental specular HREELS spectrum of the PtPc multilayer on Ag($111$) and (b) the calculated spectrum of the isolated molecule are shown. The calculated vibrational frequencies of the out-of-plane $\mathrm{A_{2u}}$ and in-plane $\mathrm{E_{u}}$ IR active modes are displayed in blue and red, respectively. The representation of the atomic displacement of the \textbf{c} mode and \textbf{g} mode is shown in (c) and (d), respectively.}
	\label{fig:multilayer}
\end{figure*}

\section{experimental details}

The experiments were performed in an ultra-high vacuum (UHV) system consisting of a preparation and an analysis chamber equipped with low electron energy diffraction (LEED) and HREELS. The pressures in the chambers were  $4\times10^{-9}$ and  $4\times10^{-10}$ mbar, respectively. The Ag(111) crystal surface was prepared by $\mathrm{Ar}^{+}$ sputtering at $1$ keV followed by annealing to $730$~K until a sharp LEED pattern was observed. Thin films of PdPc and PtPc were prepared in UHV by depositing the molecules, evaporated from a home-made Knudsen cell, on the crystal kept at room temperature. The sublimation temperatures were 770~K and 670~K for PdPc and PtPc, respectively. A multilayer phase, which does not exhibit a LEED pattern, is prepared by sublimating PtPc more than ten times longer than the sublimation time necessary to obtain the ordered monolayer phase. The pressure in the preparation chamber did not exceed $1\times10^{-8}$ mbar during sublimation. 

After the layer was prepared, its long range order was checked by LEED. Vibrational features were recorded with a primary electron beam energy of 6.3 eV in both specular and off-specular geometries. The incoming beam is directed to the sample with an angle of $45^\circ$ with respect to surface normal. In the specular (off-specular) geometry the electrons are detected at an angle of $45^\circ$ ($60^\circ$) with respect to the surface normal. The energy resolution, estimated from the full width at half maximum (FWHM) of the elastic peak, is in the range of $16-22~\mathrm{cm}^{-1}$ ($2-2.7~\mathrm{meV}$). In order to interpret the complex vibrational spectra, we performed theoretical calculations of isolated PtPc and PdPc molecules using \textsc{Gaussian} \cite{g09}. The DFT calculations of their electronic structure and of their vibrational eigenfrequencies were carried out using the LanL2DZ basis set and the B3LYP functional. The theoretical vibration energies were compressed by a factor $0.9456$ \cite{Liu2007, Maslov2006}. 

\section{Platinum-phthalocyanine on Ag(111)}

\subsection{Long range order}

PtPc molecules form ordered two-dimensional structures on the Ag(111) surface. Like for most of the MPc molecules, two diffuse rings are detected at room temperature in LEED at low coverage after annealing a thick layer up to 730~K (see Ref.~\onlinecite{SM}). This two-dimensional disordered phase has already been observed in the case of, for example, CuPc on Ag(111) and Au(111) \cite{Kroger2010, Stadtmuller2011, Schonauer2016}. Stadler \textit{et al.} interpreted the presence of this two-dimensional gas phase at low coverage as the result of an intermolecular repulsion \cite{Stadler2009a}. As the density of the PtPc molecules on the surface increases, diffraction spots are detected in LEED at room temperature (ordered phase). According to high resolution structural measurements for CuPc, the unit cell formed by the CuPc molecules decreases continuously in size as the coverage is increased. Our LEED measurements indicate a similar behavior for PtPc on Ag(111). 

\subsection{Vibrational properties of the multilayer and mode assignment}

\begin{table*}
\centering
\begin{tabular}{lccccrl}   
\hline 
\hline 
Modes &  Multilayer&  Ordered monolayer&  DFT &  Symmetry &  \multicolumn{2}{c}{Description} \tabularnewline 
\hline
\textbf{a} &  $145$  &145&  $136$         &  $\mathrm{A_{2u}}$ & OP bend:& Molecule       \tabularnewline
\textbf{R1}&  -   &255&      -          &  	-		& -									\tabularnewline
\textbf{b} &  $310$  &-&  $301$         &  $\mathrm{A_{2u}}$ & OP bend:& Molecule             \tabularnewline
\textbf{c} &  $380$  &350&  $382$         &  $\mathrm{A_{2u}}$ & OP bend:& N atoms                     \tabularnewline
\textbf{d} &  $430$  &430&  $434$         &  $\mathrm{A_{2u}}$ & OP bend:& N + C atoms                      \tabularnewline
\textbf{e} &  $505$  &-&  $490$         &  $\mathrm{E_{u}}$  & IP sciss:& N + C atoms                           \tabularnewline
\textbf{R2}&-     &565  &-               & -& -												\tabularnewline
\textbf{f} &  $575$  &-&  $574$         &  $\mathrm{E_{u}}$  & IP deform: &Molecule                                      \tabularnewline
\textbf{R3}&-     &655&-						&   - & -								\tabularnewline
\textbf{g} &  $730$  &715&  $738$         &  $\mathrm{A_{2u}}$ & OP bend: &Main ring                  \tabularnewline
\textbf{h} &  $765$  &765&  $770$         &  $\mathrm{A_{2u}}$ & OP bend:& N atoms + Phenyl groups       \tabularnewline
\textbf{i} &  $880$  &-&  $869$         &  $\mathrm{E_{u}}$  & IP deform: &Molecule                                          \tabularnewline
\textbf{j} &  $925$  &-&  $956$         &  $\mathrm{A_{2u}}$ & OP bend: &Phenyl groups                  \tabularnewline
\textbf{k} &  $1075-1200$ &-&  $1031-1159$   &  $\mathrm{E_{u}}$  & IP deform/stretch:& Molecule         \tabularnewline
\textbf{F1} &  -  &$1075-1200$&  -         &  -      &-            \tabularnewline
\textbf{l} &  $1285-1375$ &-&  $1260-1323$   &  $\mathrm{E_{u}}$  & IP rock/stretch: &N atoms + Phenyl groups \tabularnewline
\textbf{F2} &  - &$1285-1375$&  -   &  -  & -                      \tabularnewline
\textbf{m} &  $1410-1530$ &-&  $1376-1465$   &  $\mathrm{E_{u}}$  & IP stretch:& Main ring                            \tabularnewline
\textbf{F3} &  - &$1410-1530$&  -   &  -&-                            \tabularnewline
\textbf{n} &  $1585-1640$ &-&  $1546-1574$        &  $\mathrm{E_{u}}$  & IP stretch: &Phenyl groups      \tabularnewline                  
\textbf{o} &  $3010-3030$ &-&  $3025-3064$   &  $\mathrm{E_{u}}$  & IP stretch: &H atoms      \tabularnewline
\hline
\hline
\end{tabular}
\caption{List of the experimental vibrational modes present in the PtPc multilayer and ordered monolayer phase specular spectra on Ag(111) accompanied by their description and symmetries. The theoretical frequencies are compressed by a factor 0.9456. The acronyms stand for: OP=out-of-plane, IP=in-plane; bend=bending mode, deform=deformation mode, sciss=scissoring mode, stretch=stretching mode, rock=rocking mode. All values are given in $\mathrm{cm}^{-1}$. } 
\label{table}
\end{table*}

PtPc has $3N-6=165$ distinct vibrational modes ($N=57$ is the number of atoms in the molecule). Its symmetry group is $\mathrm{D_{4h}}$, its vibrational representation is $\Gamma=14 \mathrm{A_{1g}} + 13\mathrm{A_{2g}} + 14 \mathrm{B_{1g}} +  14 \mathrm{B_{2g}} + 26 \mathrm{E_{g}} + 56 \mathrm{E_{u}} + 8 \mathrm{A_{2u}} + 6 \mathrm{A_{1u}} + 7 \mathrm{B_{1u}} +7 \mathrm{B_{2u}}$. Modes transforming according to the irreducible representations $\mathrm{A_{2u}}$ and $\mathrm{E_{u}}$ are IR active modes, i.e.,~the motion of the atoms produces a dynamic dipole moment $\vec{\mu}_\mathrm{dyn}$. Depending on the direction of $\vec{\mu}_\mathrm{dyn}$, the IR active modes can be classified as in-plane modes, in which $\vec{\mu}_\mathrm{dyn}$ lies in the molecular $xy$-plane ($\mathrm{E_{u}}(x,y)$ modes), and as out-of-plane modes, in which $\vec{\mu}_\mathrm{dyn}$ is oriented perpendicular to the molecular plane along the $z$-direction ($\mathrm{A_{2u}}(z)$ modes). The $\mathrm{A_{1g}}$, $\mathrm{E_{g}}$, $\mathrm{B_{1g}}$ and $\mathrm{B_{2g}}$ modes are Raman (R) active modes, i.e.,~the atomic vibration induces a change of the molecular polarizability.  

In this work we are interested in IR active modes, because HREELS is sensitive to the dynamic dipole moment $\vec{\mu}_\mathrm{dyn}$. According to the surface selection rule \cite{Ibach1982}, modes for which $\vec{\mu}_\mathrm{dyn}$ is oriented perpendicular to the surface are excited in the dipole scattering mechanism (specular geometry), whereas modes having a dynamic dipole oriented parallel to the surface are suppressed, because the $\vec{\mu}_\mathrm{dyn}^*$ produced by the image charges in the substrate is anti-parallel to $\vec{\mu}_\mathrm{dyn}$, leading to a cancellation of the two. In the language of group theory, the surface selection rule states that only totally symmetric modes ($\mathrm{A_{1}}$, $\mathrm{A'}$ and $\mathrm{A}$ representations) of the molecule-substrate complex may be visible in HREELS. In contrast, all modes can be excited in the impact scattering regime (off-specular geometry). The appearance and/or disappearance of specific modes in the HREELS spectrum, depending on the measurement geometry, give important information on the symmetry of the molecule-substrate complex. Therefore, an accurate assignment of the vibrational features is necessary. 

The assignment is carried out by comparing the experimental peaks in the spectrum of a multilayer with calculated frequencies of the isolated PtPc molecule. In the multilayer, the effect of the substrate is reduced and thus the spectral properties are expected to be similar to the isolated molecule. Fig.~\ref{fig:multilayer}(a) shows the experimental spectrum of a PtPc multilayer on Ag(111), acquired in specular geometry, in comparison with the calculated spectrum of the isolated PtPc molecule (Fig.~\ref{fig:multilayer}(b)). The calculated spectrum contains a gaussian broadening of $20~\mathrm{cm}^{-1}$, similar to the experimentally observed one, and is composed of $\mathrm{A_{2u}}$ (blue lines) and $\mathrm{E_{u}}$ (red lines) modes only, because these are the sole IR active vibrations. We note that all experimental features can be identified in good agreement with theory. A simplified description of the modes is given in Tab.~\ref{table} (see Ref.~\onlinecite{SM} for details). The spectrum is dominated by the $\mathrm{A_{2u}}$ modes, with dynamic dipole moments perpendicular to the molecular plane, such as the strongest vibrations \textbf{a} at $145~\mathrm{cm}^{-1}$ (out-of-plane bending of the whole molecule except Pt) and \textbf{g} at $730~\mathrm{cm}^{-1}$ (out-of-plane bending of the central ring of alternating C and N atoms around the metal porphyrazin macrocycle together with the H atoms, Fig.~\ref{fig:multilayer}(d)). Compared to experiment, the theoretical $\mathrm{E_{u}}$ mode intensities appear reduced, suggesting a predominantly flat adsorption orientation of the molecules in the multilayer. However, the intensity ratio of the in-plane modes ($\mathrm{E_{u}}$) and out-of-plane modes ($\mathrm{A_{2u}}$) is conserved in the off-specular spectrum (not shown). This can be caused by a residual contribution of the dipole scattering in the off-specular geometry due to, e.g., surface roughness.    

\subsection{Vibrational Properties of the ordered monolayer phase and molecular symmetry reduction}
\label{first}

\begin{figure*}
	\centering
		\includegraphics [width=\textwidth]{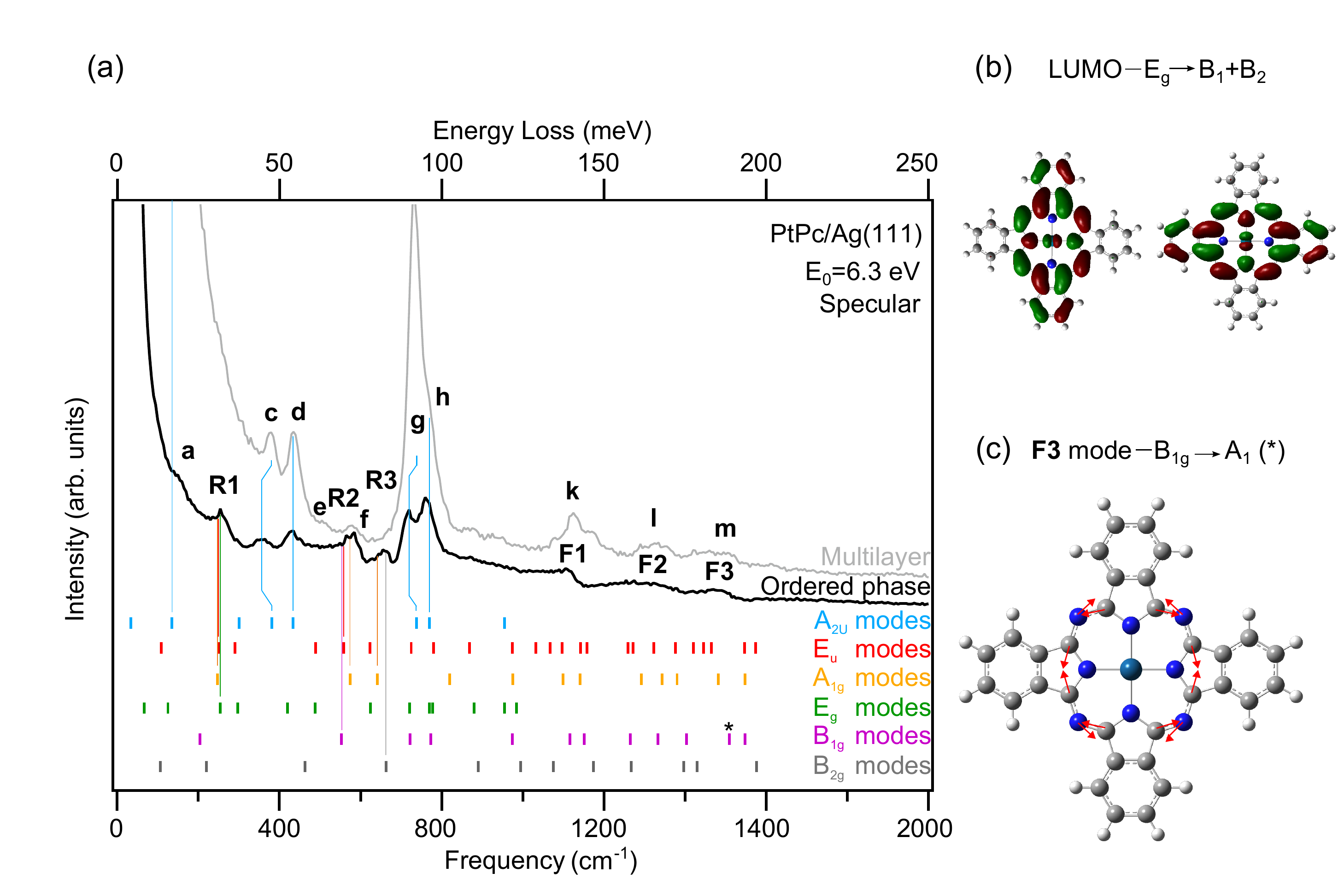}
		\caption{(a) The comparison of the specular HREELS spectra of the PtPc ordered monolayer phase and multilayer on Ag(111). The simulated vibrational modes of the isolated molecule are indicated as color bars. Note that the vibrational properties of the ordered monolayer phase do not change with the coverage. (b) The calculated charge density of the two degenerate LUMOs of isolated PtPc. (c) Representation of the Raman mode of $\mathrm{B_{1g}}$ symmetry at $1509~\mathrm{cm}^{-1}$ (marked with * in panel (a)) that is used to the fit the Fano peak \textbf{F3} at  approximately $1500~\mathrm{cm}^{-1}$ (see text).}
	\label{fig:comparison}
\end{figure*}

The vibrational properties of the ordered monolayer phase are discussed in comparison with those of the multilayer. Fig.~\ref{fig:comparison}(a) shows specular HREELS spectra of the ordered monolayer phase and the multilayer. In the ordered monolayer, most of the $\mathrm{E_{u}}$ in-plane modes are absent, especially between 1000 and $1600~\mathrm{cm}^{-1}$. But in this range three asymmetric peaks (\textbf{F1}, \textbf{F2} and \textbf{F3}) are detected. Their origin will be discussed later. The absence of the in-plane vibrations indicates that the surface selection rule strongly attenuates the in-plane $\mathrm{E_{u}}$ modes in comparison to the out-of-plane $\mathrm{A_{2u}}$ modes. We conclude that the molecules in the monolayer are oriented strictly parallel to the Ag surface. 

\subsubsection{Out-of-plane $\mathrm{A_{2u}}$ modes}

Like in the multilayer, all of the out-of-plane $\mathrm{A_{2u}}$ modes are detected in the spectrum of the ordered monolayer phase. However, in the latter the \textbf{c} and \textbf{g} modes are shifted to lower wavenumbers. The shift of the \textbf{c} mode ($\simeq30~\mathrm{cm}^{-1}$) is larger than that of the \textbf{g} mode ($\simeq15~\mathrm{cm}^{-1}$). The \textbf{c} mode is associated with the out-of-plane bending of the aza bridge N atoms against the pyrrole N atoms (Fig.~\ref{fig:multilayer}(c)), while in the \textbf{g} mode all N atoms move in-phase against the C atoms in the porphyrazin macrocycle and the H atoms in the phenyl groups (Fig.~\ref{fig:multilayer}(d)). We assign the sizable shifts of these two modes to the molecule-substrate interaction. In contrast to FePc and CoPc on Ag(111) \cite{Ohta2014, Baran2013}, for which the central metal forms a covalent bond with the substrate, the molecule-substrate interaction in PtPc takes place through the porphyrazin macrocycle, because neither the \textbf{c} nor the \textbf{g} modes, which are the ones that are most strongly influenced by the substrate, involve the central metal atom, unlike the \textbf{a} mode for example, which involves the metal atom but does not show an appreciable shift. This observation is in agreement with the general trend of a reduced metal participation in the molecule-substrate bond as the number of $d$-electrons increases \cite{Gottfried2015}. The observation that the shift of \textbf{c} is larger than that of \textbf{g}, in conjunction with the fact that the \textbf{c} mode displacement is due to the N atoms only while in the \textbf{g} mode the displacements of N atoms are  relatively small, suggests that most of the interaction between the molecule and the Ag substrate occurs via the N atoms. 

\begin{table}
\centering
\begin{tabular}{llll} 
\hline
\hline
 $\mathrm{D_{4h}}$&$\mathrm{C_{4v}}$&$\mathrm{C_{2v}}(\sigma_v)$&$\mathrm{C_{2v}}(\sigma_d)$ \tabularnewline 
\hline  
$\mathrm{E_{u}}(x,y)$&$\mathrm{E}(x,y)$&$\mathrm{B_{1}}(x)+\mathrm{B_{2}}(y)$&$\mathrm{B_{1}}(x)+\mathrm{B_{2}}(y)$      \tabularnewline
$\mathrm{A_{1g}}$&$\mathrm{A_1}(z)$&$\mathrm{A_1}(z)$&$\mathrm{A_1}(z)$   \tabularnewline
$\mathrm{E_{g}}$&$\mathrm{E}(x,y)$&$\mathrm{B_{1}}(x)+\mathrm{B_{2}}(y)$&$\mathrm{B_{1}}(x)+\mathrm{B_{2}}(y)$		\tabularnewline
$\mathrm{B_{2g}}$&$\mathrm{B_{2}}$&$\mathrm{A_2}$&$\mathrm{A_1}(z)$   \tabularnewline
$\mathrm{B_{1g}}$&$\mathrm{B_{1}}$&$\mathrm{A_1}(z)$&$\mathrm{A_2}$   \tabularnewline
\hline
\hline
\end{tabular}
\caption{Correlation tables of the $\mathrm{D_{4h}}$ symmetry group limited to the possible modes assignment of the \textbf{R} and \textbf{F} modes. The notation $(x,y)$ and $(z)$ define the in-plane and out-of-plane character of the representation. The other representations are not IR active.} 
\label{correlation}
\end{table}

\begin{table}
\centering
\begin{tabular}{ll} 
\hline
\hline 
Mode& Symmetry \tabularnewline
\hline
\textbf{R1} &$\mathrm{A_{1g}}$/$\mathrm{E_{u}}(x,y)$/$\mathrm{E_{g}}$      \tabularnewline
\textbf{R2} &$\mathrm{A_{1g}}$/$\mathrm{E_{u}}(x,y)$/$\mathrm{B_{1g}}$        \tabularnewline
\textbf{R3} &$\mathrm{A_{1g}}$/$\mathrm{B_{2g}}$     \tabularnewline
\hline
\textbf{F1} &$\mathrm{A_{1g}}$/$\mathrm{E_{u}}(x,y)$/$\mathrm{B_{1g}}$     \tabularnewline
\textbf{F2} &$\mathrm{A_{1g}}$/$\mathrm{E_{u}}(x,y)$/$\mathrm{B_{1g}}$/$\mathrm{B_{2g}}$     \tabularnewline
\textbf{F3} &$\mathrm{B_{1g}}$     \tabularnewline
\hline
\hline
\end{tabular}
\caption{Possible assignments of the \textbf{R} and \textbf{F} modes observed in the ordered monolayer phase of PtPc on Ag(111). The \textbf{F} modes assignments are based on the fit results presented in Ref.~\onlinecite{SM}. } 
\label{modes}
\end{table}

\subsubsection{Raman modes}
\label{raman modes}

Another noteworthy difference between the vibrational properties of the ordered monolayer phase and the multilayer is the appearance of new modes labeled \textbf{R1}, \textbf{R2} and \textbf{R3} in Fig.~\ref{fig:comparison}(a). Based on their frequencies, the \textbf{R1} peak at $255~\mathrm{cm}^{-1}$ can be interpreted either as a $\mathrm{E_{u}}$, $\mathrm{A_{1g}}$ or as a $\mathrm{E_{g}}$ mode, while the \textbf{R2} peak at $565~\mathrm{cm}^{-1}$ may be attributed to either a $\mathrm{E_{u}}$, $\mathrm{A_{1g}}$ or a $\mathrm{B_{1g}}$ mode. Finally, the \textbf{R3} peak at $655~\mathrm{cm}^{-1}$ can either be ascribed to a $\mathrm{A_{1g}}$ or a $\mathrm{B_{2g}}$ mode. The possible assignments are summarized in Tab.~\ref{modes}. Ruling out the $\mathrm{E_{u}}$ modes, because as in-plane IR modes they should be screened by the metal surface, only the $\mathrm{A_{1g}}$, $\mathrm{E_{g}}$, $\mathrm{B_{1g}}$ and $\mathrm{B_{2g}}$ modes remain as possible assignments. However, they are R active modes and as such not expected to contribute to the spectrum of an isolated molecule. But they can be activated if the molecular symmetry is reduced. Thus, a symmetry reduction of the molecule in contact with the surface must be considered.

When a PtPc molecule adsorbs with its plane parallel to the surface, its molecular symmetry is reduced from the original $\mathrm{D_{4h}}$ group. Specifically, at least the in-surface-plane $(x,y)$ mirror symmetry is broken. A PtPc molecule lacking its $(x,y)$-plane symmetry belongs to the $\mathrm{C_{4v}}$ group. However, further symmetry reductions are possible, e.g., to the $\mathrm{C_{2v}}$ group, if the fourfold symmetry of the molecule is broken. With the help of so-called correlation tables it is possible to deduce how the modes (representations) of the isolated molecule ($\mathrm{D_{4h}}$) relate to the ones of the molecule-substrate system of reduced symmetry ($\mathrm{C_{4v}}$ or lower) \cite{Ibach1982}. An adaptation of the correlation tables of the $\mathrm{D_{4h}}$ group from Ref.~\onlinecite{Ibach1982}, considering the observed \textbf{R} modes (and the \textbf{F} modes), is given in Tab.~\ref{correlation}. 

In principle, there is a chance that the observation of the R modes allows us to deduce the symmetry of the molecule-substrate complex. For example, a $\mathrm{B_{2g}}$ mode would become activated for HREELS dipole scattering only if the $\mathrm{D_{4h}}$ symmetry was reduced to $\mathrm{C_{2v}(\sigma_{d})}$ upon adsorption (Tab.~\ref{correlation}). Hence, we must look for a single reduced symmetry group in which the vibrational modes of the free molecule which we assign to \textbf{R1}, \textbf{R2} and \textbf{R3} map onto a totally symmetric representation ($\mathrm{A_{1}}$, $\mathrm{A'}$ or $\mathrm{A}$). However, since it is a possibility that \textbf{R1}, \textbf{R2} and \textbf{R3} could all originate from $\mathrm{A_{1g}}$ modes (see Tab.~\ref{modes}), all three would in this case be visible in dipole scattering HREELS irrespective of the symmetry of the molecule-substrate complex ($\mathrm{C_{4v}}$, $\mathrm{C_{2v}({\sigma_v})}$ or $\mathrm{C_{2v}({\sigma_d})}$). Hence, it is not possible to conclude on the basis of \textbf{R1}, \textbf{R2} and \textbf{R3} alone whether PtPc preserves its fourfold symmetry upon adsorption ($\mathrm{C_{4v}}$) or lowers its symmetry to a twofold one ($\mathrm{C_{2v}({\sigma_v})}$/$\mathrm{C_{2v}({\sigma_d})}$).

\begin{figure}
	\centering
		\includegraphics [width=\columnwidth]{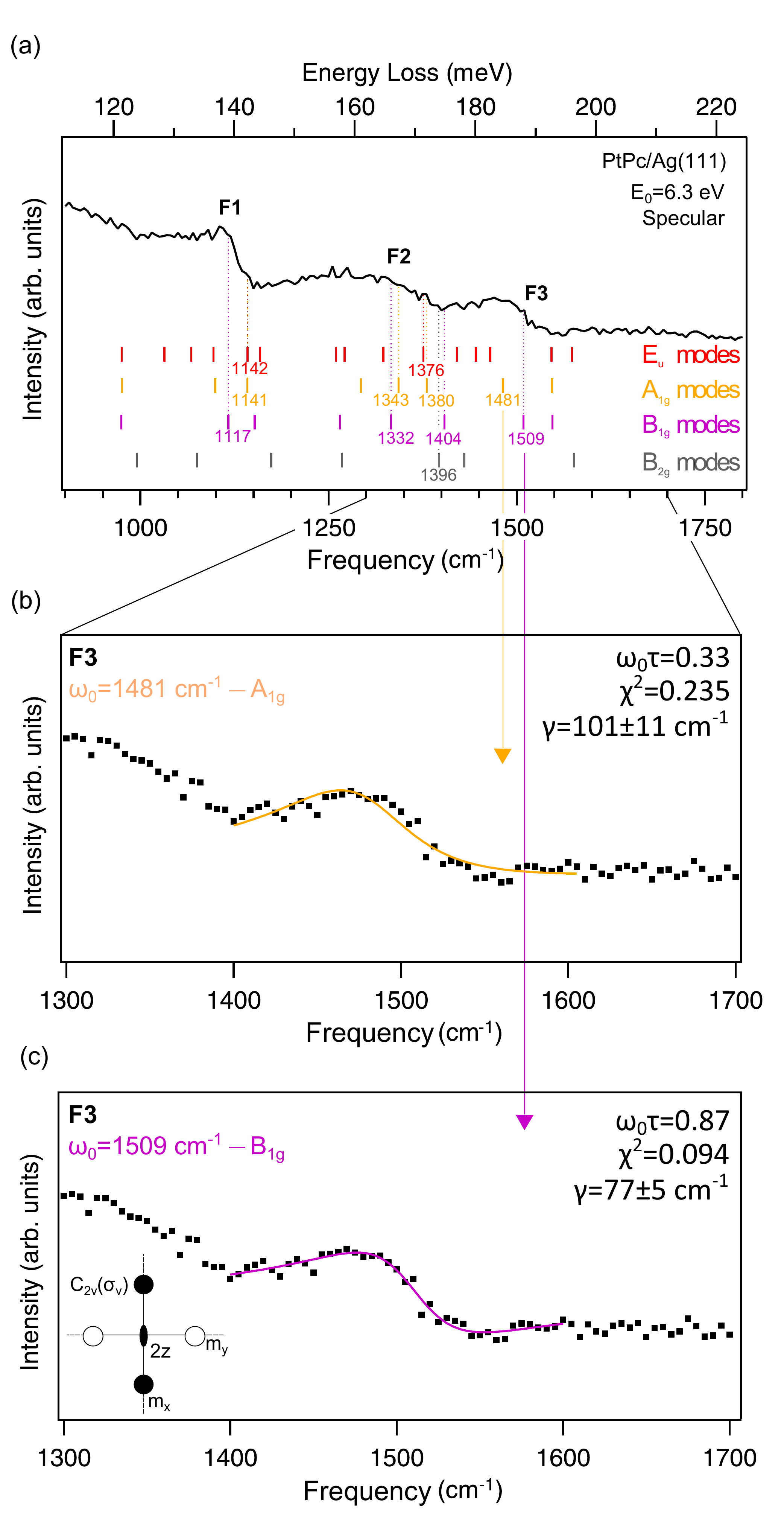}
		\caption{(a) Zoom of the ordered monolayer phase spectrum of PtPc on Ag(111) showing the asymmetric peaks \textbf{F1}, \textbf{F2} and \textbf{F3} together with the theoretical vibrational modes (vertical bars). Each peak has been fitted using all the theoretical vibrations visible in panel (a), and all the $\omega_{0}$ values which give reasonable fits of \textbf{F1}, \textbf{F2} and \textbf{F3} are shown with dotted lines. An example of bad fit of \textbf{F3} is shown in panel (a) using $\omega_0=1481~ \mathrm{cm}^{-1}$, and the best fit of \textbf{F3} is shown in panel (c) using $\omega_0=1509~\mathrm{cm}^{-1}$. The symmetry ($\mathrm{C_{2v}}(\sigma_v)$) of the PtPc molecule in contact with Ag(111) is depicted in the inset of (c). }
	\label{fig:fano}
\end{figure}

\subsubsection{Asymmetric peaks}
\label{Asymmetric peaks}

In order to identify the symmetry of the PtPc molecule on the Ag surface, we now turn to the analysis of the \textbf{F} modes. Between $1000$ and $1600~\mathrm{cm}^{-1}$, where the $\mathrm{E_{u}}$ modes are excited in the multilayer spectrum (Fig.~\ref{fig:multilayer}(a)), three asymmetric features, labeled \textbf{F1}, \textbf{F2} and \textbf{F3} in Fig.~\ref{fig:comparison}(a), are visible in the spectrum of the ordered monolayer. Asymmetric line shapes of Fano type are well known for molecular and atomic adsorbates on metallic surfaces in the presence of an IDCT \cite{ Chabal1985, Kroger1997, Tautz2002, Amsalem2009, Duo2010, Kroeger2011, Braatz2012, Rosenow2016}. A prerequisite for the occurrence of IDCT is the partial filling of an orbital that couples to a molecular vibration. Then, as the vibration is excited, the associated distortion of the molecule along the vibrational coordinate modulates the orbital energy periodically, leading to an oscillatory charge flow between the molecule and the substrate that is effectively pumped by the vibration. In the presence of IDCT, the dynamical dipole moment $\mu_\mathrm{dyn}$ of the vibration is the sum of the ionic and the electronic contributions, $\mu_\mathrm{ion}$ and $\mu_\mathrm{IDCT}$. If the dynamic electron transfer between the orbital and the substrate, yielding  $\mu_\mathrm{IDCT}$, is not in phase with the ion movement, leading to $\mu_\mathrm{ion}$, i.e. if the Born-Oppenheimer approximation breaks down, $\mu_\mathrm{dyn}$ acquires an imaginary part ($\mu_\mathrm{dyn} = \mu_1 + i\mu_2$). This leads to a damping of the adsorbate vibration and yields an asymmetric Fano line shape. By introducing the asymmetry parameter $\omega_0\tau=\mu_2/\mu_1$ the resulting line shape can be expressed as \cite{Langreth1985, Crljen1987,Kroger2006}

\begin{equation}
L(\omega)= a+c\gamma\omega\omega_{0} \frac{[1-(\frac{\tau}{\gamma})(\omega^{2}-\omega_{0}^{2})]^{2}}{(\omega^{2}-\omega_{0}^{2})^{2} + (\gamma\omega)^{2}},
\label{fano}
\end{equation}
where $a$ and $c$ are constants, and $\gamma$ denotes the width of the Fano profile. $\omega_0$ is the vibrational frequency of a given mode and $\tau$, the asymmetry ratio, is (negative) positive if $\mu_2$ is (anti-) parallel to $\mu_1$.  

In order to identify which vibrational modes are involved in the IDCT that produces \textbf{F1}, \textbf{F2} and \textbf{F3}, Eq.~1 is used to fit these features, using values of $\omega_0$ that are determined by DFT calculations of an isolated molecule. All the calculated $\omega_0$ frequencies in the range from $900$ to $1800~\mathrm{cm}^{-1}$ have been tested (Fig.~\ref{fig:fano}(a)). The fits can be found in the supplement~\cite{SM}. The few modes that yield reasonable fits for \textbf{F1} and \textbf{F2} are summarized in Tab.~\ref{modes}. Because both features may possibly be assigned to a $\mathrm{A_{1g}}$ mode, whose representation in the $\mathrm{D_{4h}}$ group transforms into a totally symmetric one in both the $\mathrm{C_{4v}}$ or $\mathrm{C_{2v}}$ groups, it is not possible to identify the symmetry of the molecule-substrate complex by considering these modes. 

Interestingly, however, \textbf{F3} at about $1500~\mathrm{cm}^{-1}$ can only be fitted well when using a $\mathrm{B_{1g}}$ vibrational mode, namely the one at $\omega_0=1509~\mathrm{cm}^{-1}$, see Fig.~\ref{fig:fano}(c). The fitted value of $\gamma$ ($77 \pm 5~\mathrm{cm}^{-1}$) compares well to the results for ZnPc/Ag(110) ($60~\mathrm{cm}^{-1}$) \cite{Amsalem2009}. For comparison, an example of a bad fit using an $\mathrm{A_{1g}}$ mode ($\omega_0=1481~\mathrm{cm}^{-1}$) is also shown in Fig.~\ref{fig:fano}(b). Since the R active $\mathrm{B_{1g}}$ mode of the $\mathrm{D_{4h}}$ symmetry group transforms into an IR active $\mathrm{A_{1}}$ mode only in the $\mathrm{C_{2v}(\sigma_v)}$ group (Tab.~\ref{correlation}), we conclude that the molecular symmetry must be reduced to twofold. The full symmetry of the PtPc molecule on the Ag(111) surface is schematically shown in the inset of Fig.~\ref{fig:fano}. 

However, the mere appearance of the $\mathrm{B_{1g}}$ mode in the spectrum does not yet clarify the origin of the symmetry reduction $\mathrm{D_{4h}}\rightarrow \mathrm{C_{2v}(\sigma_v)}$. In principle, it is possible that because of a geometric distortion of the molecule $\mathrm{\mu_{ion}}$ is tilted out of the surface plane and acquires a perpendicular contribution. But in this case $\mathrm{\mu_{dyn}}$ would necessarily remain real, and we would observe a conventional, symmetric line shape for \textbf{F3}. This is evidently not the case. Its Fano line shape shows that $\mathrm{\mu_{dyn}}$ of \textbf{F3} is complex. Hence, there must be a phase-shifted electronic contribution $\mathrm{\mu_{IDCT}}$ to its overall dynamical dipole moment $\mathrm{\mu_{dyn}}$. 

We have already mentioned above that for PtPc/Ag(111) the LUMO is the most likely partner in IDCT, because it becomes filled upon adsorption and straddles the Fermi energy. Moreover, a comparison of the elongation pattern of \textbf{F3}, which predominantly involves the stretching of the C-N bonds in the porphyrazine macrocycle (Fig.~\ref{fig:comparison}(c)), with the lobular structure of the LUMO (Fig.~\ref{fig:comparison}(b)) shows that the ionic motion of \textbf{F3} should indeed couple well to the LUMO. Comparing Figs.~\ref{fig:comparison}(b) and (c) moreover shows that exciting \textbf{F3} with positive amplitude has the same effect on the $\mathrm{B_1}$-LUMO as exciting \textbf{F3} with negative amplitude has on the $\mathrm{B_2}$-LUMO. Hence, if $\mathrm{B_1}$ and $\mathrm{B_2}$ were still degenerate, dynamic filling and depletion of the two would be 180$^\circ$ out of phase (assuming that the phase lag relative to the ionic movement is the same for both LUMOs, which is expected to be a good approximation, in spite of the  observed symmetry breaking, see below). In other words, in one half of the vibrational oscillation cycle the $\mathrm{B_1}$-LUMO would be filled, while the $\mathrm{B_2}$-LUMO would be depleted, and vice versa in the other half. Overall, no $\mathrm{\mu_{IDCT}}$ would prevail (small differences between the IDCT in the $\mathrm{B_1}$- and $\mathrm{B_2}$-LUMOs notwithstanding).

However, in the experimental spectrum a substantial $\mathrm{\mu_{IDCT}}$ is observed, as argued above. This can mean two things. Either the two LUMOs are still degenerate and their dynamic couplings to the \textbf{F3} vibration are different, or the LUMO levels $\mathrm{B_1}$ and $\mathrm{B_2}$ are indeed split. We consider it unlikely that the static level positions of the LUMOs are not affected by symmetry breaking, while their dynamic couplings to \textbf{F3} are. Therefore, we conclude that the $\mathrm{B_1}$- and $\mathrm{B_2}$-LUMO levels must be split, i.e.~the degeneracy of the PtPc LUMO is lost upon adsorption on Ag(111). One may speculate that this LUMO splitting proceeds via a geometric distortion that is induced by the environment (the Ag(111) surface),  enhanced by a Jahn-Teller-like internal stabilization of this external geometric distortion via the splitting of the electronic LUMO state\cite{Jahn1937,Kopper2009,Uhlmann2013}.          

\begin{figure*}
	\centering
		\includegraphics [scale=0.6]{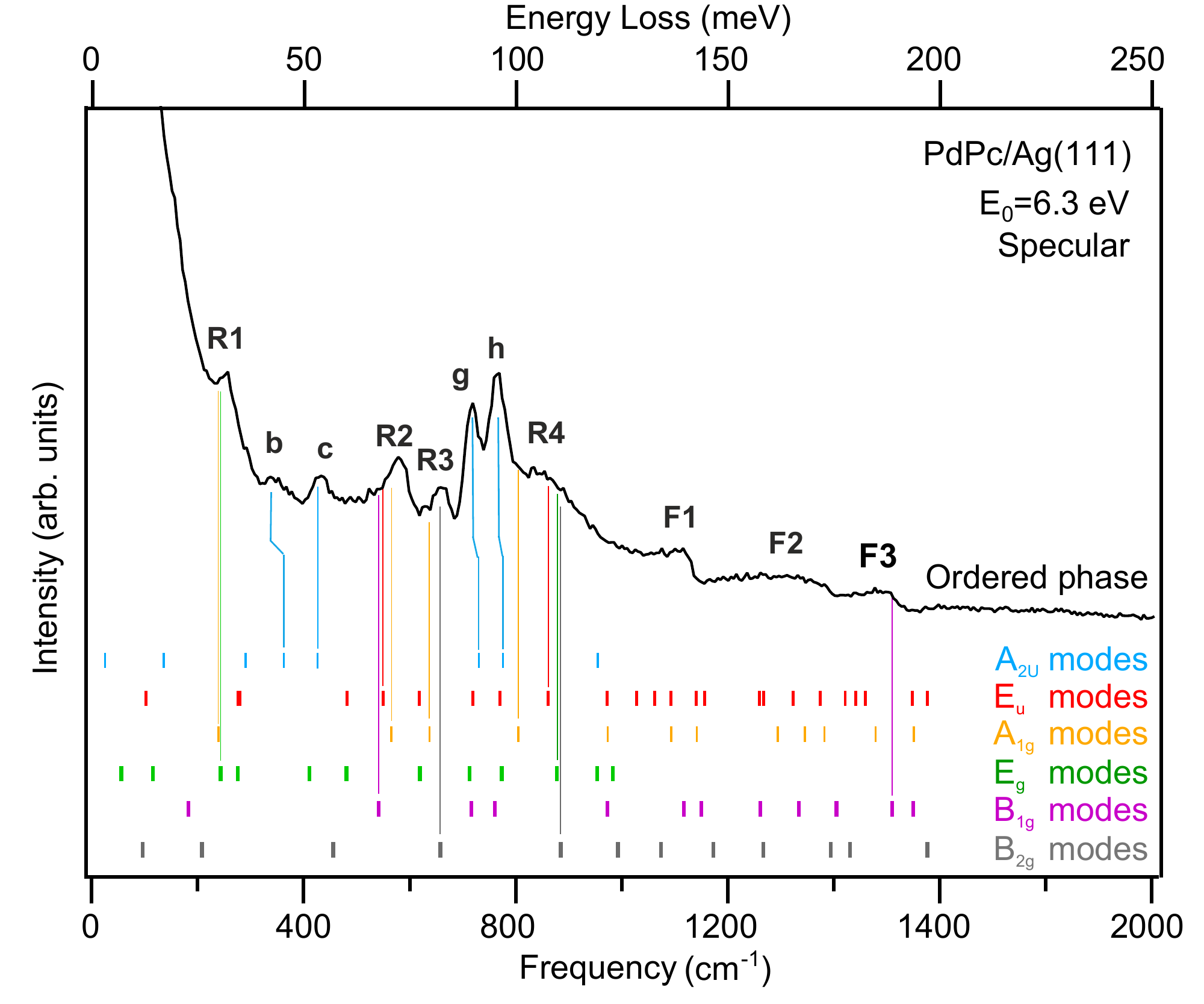}
		\caption{Specular HREELS spectra of the PdPc ordered monolayer phase on Ag(111). The simulated vibrational modes of the isolated molecule are shown as colored vertical bars.}
	\label{fig:comparisonPd}
\end{figure*}

\begin{figure}
	\centering
		\includegraphics [width=\columnwidth]{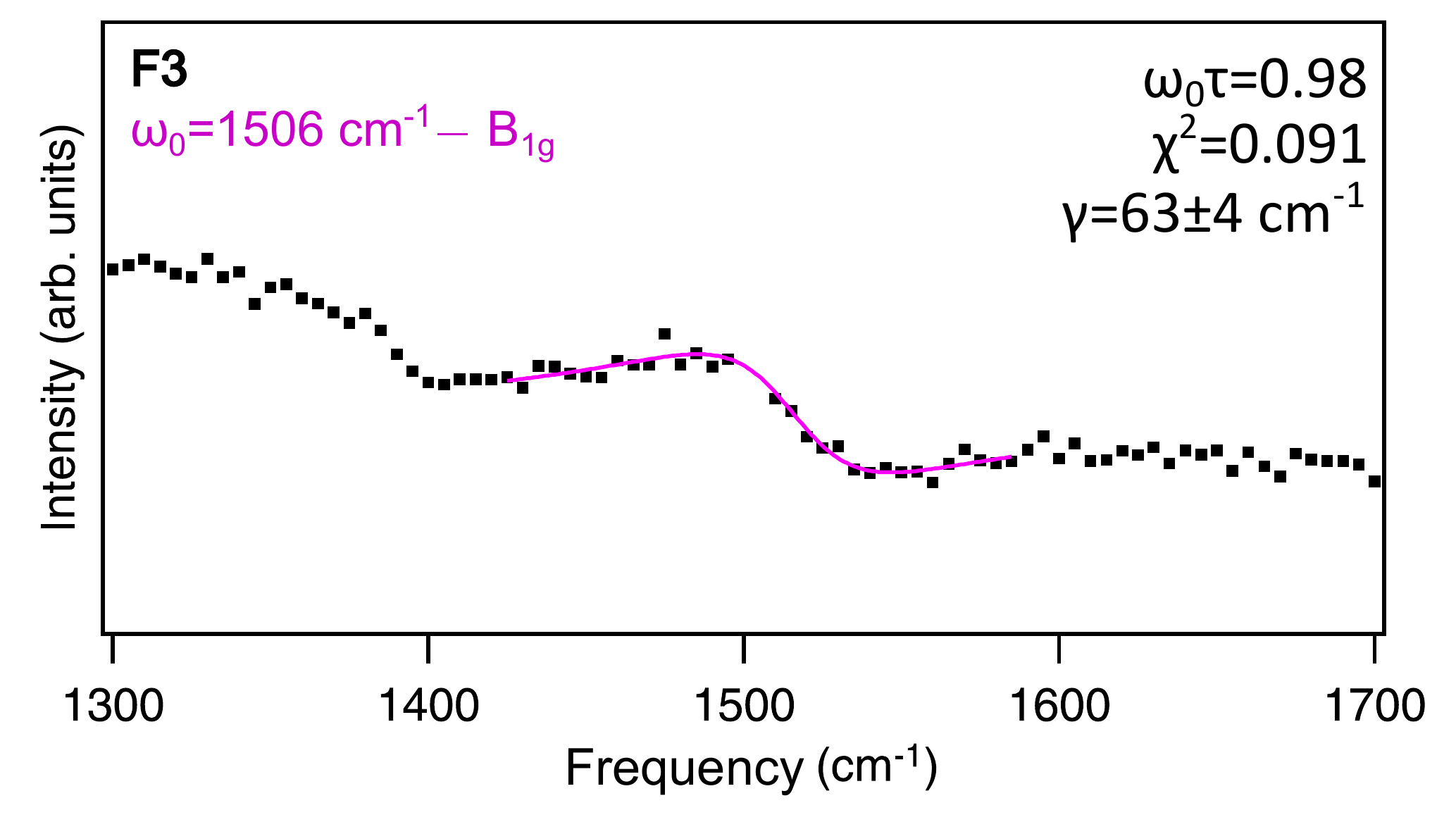}
		\caption{ Best fit of \textbf{F3} using $\omega_0=1506~\mathrm{cm}^{-1}$, for the ordered monolayer PdPc phase on Ag(111). }
	\label{fig:fanoPd}
\end{figure}

\section{Palladium-phthalocyanine on Ag(111)}
PdPc molecules adsorbed on Ag(111) show similar long range order as PtPc/Ag(111). A two-dimensional gas phase is found after annealing the thick layer up to $730$K and an ordered phase is detected in LEED for lower annealing temperatures, that is at higher coverages.

In Fig.~\ref{fig:comparisonPd} the specular HREELS spectrum of the PdPc ordered monolayer phase is shown together with the simulated results for the isolated molecule. The assignment of the experimental vibrational frequencies can be found in the supplement~\cite{SM}. As in the case of PtPc, the in-plane $\mathrm{E_{u}}$ modes are absent and only the dipole scattering allowed out-of-plane $\mathrm{A_{2u}}$ modes of the isolated molecule are present. This indicates an essentially flat adsorption geometry. The \textbf{c}, \textbf{g} and \textbf{h} modes are shifted to lower frequencies by approximately 45, 15 and $15~\mathrm{cm}^{-1}$ with respect to theory, respectively. The larger shift of the \textbf{c} mode for PdPc ($45~\mathrm{cm}^{-1}$) than for PtPc ($30~\mathrm{cm}^{-1}$) suggests a stronger interaction of PdPc with the Ag(111) substrate.  

The presence of R modes in Fig.~\ref{fig:comparisonPd} suggests that a reduction of the molecular symmetry upon adsorption. In addition to the \textbf{R1}, \textbf{R2} and \textbf{R3} peaks that are also observed for PtPc/Ag(111), the PdPc spectrum shows a peak at $850~\mathrm{cm}^{-1}$, labeled \textbf{R4}. Possible assignments of \textbf{R1}, \textbf{R2}, \textbf{R3} and \textbf{R4} are listed in Tab.~\ref{modesPd}. 
\begin{table}
\centering
\begin{tabular}{ll} 
\hline
\hline 
Mode& Symmetry \tabularnewline
\hline
\textbf{R1} &$\mathrm{A_{1g}}$/$\mathrm{E_{g}}$     \tabularnewline
\textbf{R2} &$\mathrm{A_{1g}}$/$\mathrm{E_{u}}(x,y)$/$\mathrm{B_{1g}}$        \tabularnewline
\textbf{R3} &$\mathrm{A_{1g}}$/$\mathrm{B_{2g}}$     \tabularnewline
\textbf{R4} &$\mathrm{A_{1g}}$/$\mathrm{E_{u}}(x,y)$/$\mathrm{B_{2g}}$/$\mathrm{E_{g}}$     \tabularnewline
\hline
\textbf{F1} &$\mathrm{A_{1g}}$/$\mathrm{E_{u}}(x,y)$/$\mathrm{B_{1g}}$     \tabularnewline
\textbf{F2} &$\mathrm{A_{1g}}$/$\mathrm{E_{u}}(x,y)$/$\mathrm{B_{1g}}$/$\mathrm{B_{2g}}$     \tabularnewline
\textbf{F3} &$\mathrm{B_{1g}}$     \tabularnewline
\hline
\hline
\end{tabular}
\caption{Possible assignments of the \textbf{R} and \textbf{F} modes observed in the ordered monolayer phase of PdPc on Ag(111). The \textbf{F} modes assignments are based on the fit results.} 
\label{modesPd}
\end{table}
There is the possibility to assign the \textbf{R1}, \textbf{R2}, \textbf{R3} and \textbf{R4} peaks to modes with a $\mathrm{A_{1g}}$ representation (since \textbf{R4} is very broad, it is rather difficult to assign it and a designation as a $\mathrm{A_{1g}}$ mode cannot be excluded); they would then be detectable in HREELS in specular geometry irrespective of the symmetry group of the molecule-substrate complex ($\mathrm{C_{2v}}$ or $\mathrm{C_{4v}}$), cf.~Tab.~\ref{correlation}. 

Three asymmetric peaks (\textbf{F1}, \textbf{F2} and \textbf{F3}) are observed between 1000 and $1600~\mathrm{cm}^{-1}$, which indicates that several vibrational modes are involved in an IDCT. Using Eq.~1, we single out vibrational frequencies given by DFT calculations for isolated PdPc ($\omega_0$) that yield the best fit of the experimental data as explain in Sec.~\ref{Asymmetric peaks}. As in the case of PtPc, several modes yield a reasonable fit for \textbf{F1} and \textbf{F2}, while \textbf{F3} can be fitted only with $\omega_0=1506~\mathrm{cm}^{-1}$, a mode which belongs to the $\mathrm{B_{1g}}$ representation of the $\mathrm{D_{4h}}$ symmetry group (see Fig.~\ref{fig:fanoPd}), and a $\mathrm{B_{1g}}$ mode can only be detected in HREELS in the specular direction if the molecular symmetry is reduced to $\mathrm{C_{2v}({\sigma_v})}$ (see Tab.~\ref{correlation}). As in the case of PtPc, this demonstrates that a preferential charge transfer occurs into one of the two LUMOs. Thus, the Ag(111) substrate lifts the degeneracy of the LUMO also in the case of PdPc.

\section{Summary and conclusion}
In conclusion, the vibrational properties of Pt- and Pd-phthalocyanine molecules on the Ag(111) surface show that both molecules adsorb with their molecular plane parallel to the surface in the ordered monolayer phase. The red shift of some of the out-of-plane modes reflects a moderate interaction between the meso-tetraazaporphin (porphyrazin) macrocycle of the molecule and the Ag substrate. The presence of Raman vibrational modes proves the lowering of the molecular symmetry from the $\mathrm{D_{4h}}$ group of the isolated molecule upon adsorption on Ag(111). The asymmetrical line shape of some of the molecular vibrational modes further demonstrates that the charge transfer to the molecule is involved in the symmetry reduction to $\mathrm{C_{2v}({\sigma_v})}$. Therefore, this study shows that HREELS is a valuable tool to determine the origin of the molecular degeneracy lifting upon adsorption on a metal surface.

\section{Acknowledgment}
F. C. Bocquet acknowledges financial support from the Initiative and Networking Fund of the Helmholtz Association, Postdoc Programme VH-PD-025. J. Sforzini thanks Prof. Dr. P. Jakob and Dr. P. Amsalem for fruitful discussions.

\bibliography{PdPtPcEELS_12}
\end {document}